\documentclass[10pt,letterpaper,twocolumn]{article} 

\usepackage{ol}
\usepackage{hyperref}
\usepackage{amsmath}

\begin{document}

\twocolumn[ 

\title{Localized vibrational modes in optically bound structures}
\author{Jack Ng and C.T. Chan}
\address{Department of Physics, Hong Kong University of Science and Technology, Clearwater Bay, Hong Kong, China}


\begin{abstract}
We show, through analytical theory and rigorous numerical calculations, that 
optical binding can organize a collection of particles into stable 
one-dimensional lattice. This lattice, as well as other optically-bound 
structures, are shown to exhibit spatially localized vibrational eigenmodes. 
The origin of localization here is distinct from the usual mechanisms such 
as disorder, defect, or nonlinearity, but is a consequence of the 
long-ranged nature of optical binding. For an array of particles trapped by 
an interference pattern, the stable configuration is often dictated by the 
external light source, but our calculation revealed that inter-particle 
optical binding forces can have a profound influence on the dynamics.
\end{abstract}


 ] 

\noindent Since its introduction many years ago,\cite{Ashkin:1970} optical 
manipulation has evolved into a major technique for manipulating small 
particles, and recently, simultaneous manipulations of multi-particles have 
been demonstrated.\cite{See:2003}$^{ }$It is known that in addition to the 
well-known one-body force such as the gradient force that depends on the
intensity profile, there is an optical binding (OB) force that couples the 
particles together.\cite{Burns:1990}$^{,}$\cite{Lin:2005} Nevertheless, 
for an extended array of particles, the nature of OB is not fully 
understood, although some theoretical efforts were devoted to small 
clusters.\cite{Lin:2005}$^{,}$\cite{Chaumet:2001} As the principles 
underlining these inter-particle forces are different from that of the 
traditional light-trapping, we expect some new and interesting applications.

In this paper, we demonstrate an interesting consequence of OB in a 
spatially extended structure bound by light: the existence of spatially 
localized VEM (vibrational eigenmodes). We illustrate the physics by 
considering a one-dimensional ``lattice'' bound by light. Wave localization 
is known to occur in defect or impurity sites of an otherwise ordered 
lattice. In solids, the ``defect'' can be impurity atoms that localize 
phonons, and in the intrinsic localized modes, the ``defect'' is derived 
from the nonlinearly excited particles.\cite{Campbell:2004} Here the 
localization occurs in the linear dynamics regime in an ordered array of 
identical particles without defect or disorder.

Optically bound structures have been investigated in a number of recent 
experiments. Stable cluster configurations had been 
realized\cite{Lin:2005}$^{,}$\cite{Tatarkova:2002}$^{,}$\cite{Singer:2003}$^{,}$\cite{Black:2003}$^{,}$\cite{Garces:2005}$^{,}$\cite{Mellor:2006} 
and vibrational motions were observed.\cite{Tatarkova:2002} In particular, 
the most commonly observed geometry is an one-dimensional array of 
particles, bound by a pair counterpropagating 
beams\cite{Tatarkova:2002}$^{,}$\cite{Singer:2003}$^{,}$\cite{Black:2003} 
or evanescent waves.\cite{Garces:2005}$^{,}$\cite{Mellor:2006}

Consider a linear chain of $N$ evenly spaced spheres in air. The particles have 
mass density $\rho $=1,050 kg/m$^{3}$, dielectric constant $\varepsilon 
$=2.53 ($\sim $polystyrene), and radii $a = \lambda / 10 = 52$ nm, so that 
they are small compare to the incident light's wavelength $\lambda $=520 nm. 
The particles are illuminated by the standing wave formed by a pair of 
counterpropagating plane waves

\begin{equation}
\label{eq1}
\mathord{\buildrel{\lower3pt\hbox{$\scriptscriptstyle\rightharpoonup$}}\over 
{E}} _{in} 
(\mathord{\buildrel{\lower3pt\hbox{$\scriptscriptstyle\rightharpoonup$}}\over 
{r}} ) = 2E_0 \cos \left( {kz} \right)\hat {x},
\end{equation}

\noindent
where $k$ is the wavenumber, and the intensity for each beam is set to be 0.01 
W/$\mu $m$^{2}$.\cite{Burns:1990}$^{,}$\cite{Lin:2005} 

To calculate the optical force acting on the particles, we employ the 
rigorous and highly accurate multiple scattering and Maxwell stress tensor 
(MS-MST) formalism,\cite{Lin:2005} which requires no approximation 
and subject only to numerical truncation errors (we use multipoles up to 
$L$=6). The optical force tends to drive small particles to the region of 
strong light intensity. For an array of $N$ evenly spaced particles aligned 
along the $z$-axis, one expects a stable one-dimensional lattice with a lattice 
constant of $\lambda $/2:

\begin{equation}
\label{eq2}
{\begin{array}{*{20}c}
 
{\mathord{\buildrel{\lower3pt\hbox{$\scriptscriptstyle\rightharpoonup$}}\over 
{R}} _n = \left( {0,0,n\lambda / 2} \right),} \hfill & {n = 1,2\ldots ,N} 
\hfill \\
\end{array} },
\end{equation}

\noindent
where 
$\mathord{\buildrel{\lower3pt\hbox{$\scriptscriptstyle\rightharpoonup$}}\over 
{R}} _n $ is the equilibrium position for the \textit{n-th} particle. Indeed, we found 
that the geometry defined in (\ref{eq2}) corresponds to a zero-force configuration 
and the configuration is proven to be stable by using linear stability 
analysis.\cite{Lin:2005}$^{,}$\cite{The:1} The longitudinal trapping 
(along the $z$-axis) is mainly provided by the gradient force of the incident 
beam, and it is further enhanced by OB.\cite{This:1} On the other hand, the 
transverse stability (on the \textit{xy}-plane) is solely induced by OB. We note that 
there are other beam configurations, other than that specified by (\ref{eq1}), that 
can stabilize a linear chain as demonstrated by recent experiments.$^{ 
}$\cite{Tatarkova:2002}$^{,}$\cite{Singer:2003}$^{,}$\cite{Mellor:2006}$^{,}$\cite{Chowdhury:1985}

\begin{figure}[htbp]
\centerline{\includegraphics[width=3.34in]{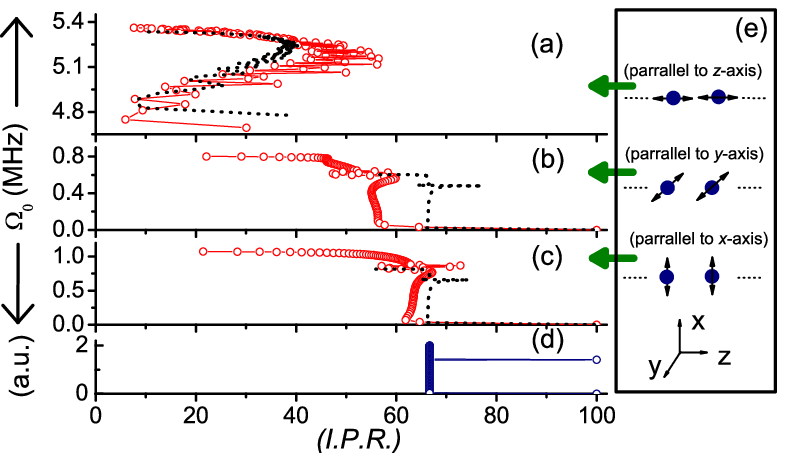}}
\caption{(Color online) Natural vibration frequencies $\Omega _{0i} $ versus the 
inverse participation ratio, for the 1D lattice with $N=100$. Panel (a), (b), 
and (c) correspond respectively to 
$\mathord{\buildrel{\lower3pt\hbox{$\scriptscriptstyle\rightharpoonup$}}\over 
{k}} $-branch, 
$(\mathord{\buildrel{\lower3pt\hbox{$\scriptscriptstyle\rightharpoonup$}}\over 
{k}} \times 
\mathord{\buildrel{\lower3pt\hbox{$\scriptscriptstyle\rightharpoonup$}}\over 
{E}} )$-branch, and 
$\mathord{\buildrel{\lower3pt\hbox{$\scriptscriptstyle\rightharpoonup$}}\over 
{E}} $-branch. The open circles are obtained by the rigorous MS-MST 
formalism and the dotted line is that of the P.E. model (\ref{eq4}). Panel (d) shows 
the $I.P.R.$ for the ``ball and spring'' model. Panel (e) shows 
schematically the direction of the particles' displacements for the three 
branches.}
\label{fig1}
\end{figure}

The VEMs are obtained by diagonalizing the force matrix 
$(\mathord{\buildrel{\lower3pt\hbox{$\scriptscriptstyle\leftrightarrow$}}\over 
{K}} )_{jk} = \partial 
(\mathord{\buildrel{\lower3pt\hbox{$\scriptscriptstyle\rightharpoonup$}}\over 
{f}} _{light} )_j / \partial (\Delta 
\mathord{\buildrel{\lower3pt\hbox{$\scriptscriptstyle\rightharpoonup$}}\over 
{x}} )_k $,\cite{Lin:2005} which is found by linearizing the optical 
force near the 
equilibrium: $\mathord{\buildrel{\lower3pt\hbox{$\scriptscriptstyle\rightharpoonup$}}\over 
{f}} _{light} \approx 
\mathord{\buildrel{\lower3pt\hbox{$\scriptscriptstyle\leftrightarrow$}}\over 
{K}} \Delta 
\mathord{\buildrel{\lower3pt\hbox{$\scriptscriptstyle\rightharpoonup$}}\over 
{x}} $, where $\Delta 
\mathord{\buildrel{\lower3pt\hbox{$\scriptscriptstyle\rightharpoonup$}}\over 
{x}} $ is the displacement vector of the \textit{i-th} particle away from its equilibrium 
configuration. The vibration profile of the VEM is described by the 
eigenvectors 
$\mathord{\buildrel{\lower3pt\hbox{$\scriptscriptstyle\rightharpoonup$}}\over 
{V}} ^{(i)}$ 
of $\mathord{\buildrel{\lower3pt\hbox{$\scriptscriptstyle\leftrightarrow$}}\over 
{K}} $, and the natural vibrational frequency is $\Omega _{0i} = ( - K_i / 
m)^{1 / 2}$ where $K_{i}$ is an eigenvalue of 
$\mathord{\buildrel{\lower3pt\hbox{$\scriptscriptstyle\leftrightarrow$}}\over 
{K}} $ and $m$ is the mass of a sphere. Due to the reflection symmetry, the 
modes fall into three separate branches (each of $N$ modes), corresponding 
to the vibrations along the three Cartesian directions (see 
Fig. \ref{fig1}(e)). We shall denote the branches as the 
$\mathord{\buildrel{\lower3pt\hbox{$\scriptscriptstyle\rightharpoonup$}}\over 
{k}} $-branch, 
$\mathord{\buildrel{\lower3pt\hbox{$\scriptscriptstyle\rightharpoonup$}}\over 
{E}} $-branch, and 
$(\mathord{\buildrel{\lower3pt\hbox{$\scriptscriptstyle\rightharpoonup$}}\over 
{k}} \times 
\mathord{\buildrel{\lower3pt\hbox{$\scriptscriptstyle\rightharpoonup$}}\over 
{E}} )$-branch, corresponding respectively to particle displacements along 
the incident 
wavevector$\mathord{\buildrel{\lower3pt\hbox{$\scriptscriptstyle\rightharpoonup$}}\over 
{k}} = \pm k\hat {z}$, the incident polarization ($x$-axis), and the $y$-axis.

The degree of localization of the modes can be quantified by calculating the 
inverse participation ratio\cite{Book:1}

\begin{equation}
\label{eq3}
\left( {I.P.R.} \right)_i = \left( {\sum\nolimits_{n = 1}^N {\left| {\left( 
{\Delta X_n^{(i)} ,\Delta Y_n^{(i)} ,\Delta Z_n^{(i)} } \right)} \right|^4} 
} \right)^{ - 1},
\end{equation}

\noindent
which indicates the number of particles participating the vibration. Here, 
the index $i$ stands for the \textit{i-th} eigenmode and $\Delta X_n^{(i)} $ is the 
vibration amplitude of the \textit{n-th} particle along the $x$-axis. A small value of 
$I.P.R.$ indicates a localized mode, while $I.P.R.\sim N$ indicates a 
delocalized mode. Fig. \ref{fig1} shows the $I.P.R.$ 
computed by the MS-MST formalism. For comparison, the $I.P.R.$ for an 
ordinary ``ball and spring'' model is also plotted in 
Fig. \ref{fig1}(d), where a lattice of 100 particles are 
connected to its nearest neighbors by a Hooke spring. As expected, the ball 
and spring model supports only propagating modes in which the displacement 
of the \textit{n-th} particle $\sim e^{inq\Delta }$, where $q$ is the phonon wavevector and 
$\Delta $ is the lattice constant. Depending on whether $q\Delta $ is an 
integer multiple of $\pi$, $I.P.R.$ takes either $\sim $200/3 or $\sim $100.

\begin{figure}[htbp]
\centerline{\includegraphics[width=3.34in]{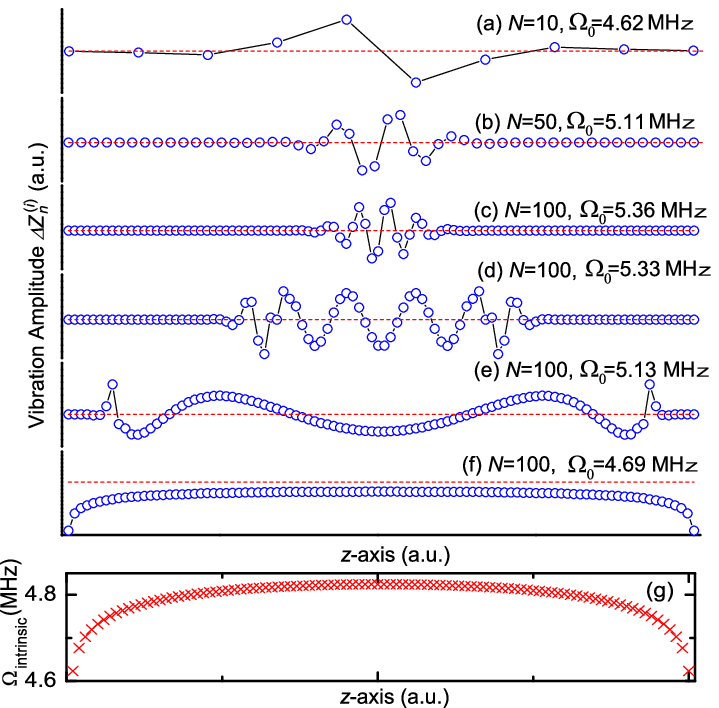}}
\caption{(Color online) The profiles of a few selected VEMs in the 
$\mathord{\buildrel{\lower3pt\hbox{$\scriptscriptstyle\rightharpoonup$}}\over 
{k}} $-branch. Dashed lines show the equilibrium positions. Each circle 
represents one particle. Panel (a)-(c): the highest frequency mode for a 
lattice containing (a): $N=10$ particles, (b): $N=50$, and (c) $N=100$. Panel 
(c)-(f): the VEMs for $N=100$, with (c) showing the highest 
frequency mode, and (d)-(e) correspond to two intermediate frequencies, and 
(f) shows the lowest frequency mode. Note that the interparticle separation 
and the size of the particles are not drawn to scale. Panel (g) shows 
$\Omega _{\mbox{intrinsic}} $ with $N=100$, see (\ref{eq9}).}
\label{fig2}
\end{figure}

In general, the VEMs of the optically-bound lattice are more localized than 
the propagating modes, especially for the 
$\mathord{\buildrel{\lower3pt\hbox{$\scriptscriptstyle\rightharpoonup$}}\over 
{k}} $-branch. A few modes selected from the 
$\mathord{\buildrel{\lower3pt\hbox{$\scriptscriptstyle\rightharpoonup$}}\over 
{k}} $-branch is shown in Fig. \ref{fig2}. The 
high-frequency modes are highly localized near the center of the lattice 
(e.g. Fig. \ref{fig2}(c)), while those with a lower 
vibrational frequency are less localized (e.g. Fig. \ref{fig2}(d)-(e)). For very low frequencies, the modes are further delocalized 
spatially (e.g. Fig. \ref{fig2}(f)), with the vibration 
being stronger on both ends. The evolution of a VEM as the number of 
particles increases is also depicted in Fig. \ref{fig2}(a)-(c); clearly the overall profile of the modes are getting more and 
more localized as the number of particle increases.

The physics of the localized mode (LM) can be captured qualitatively by a 
simple potential energy model (P.E. model)$_{.}$\cite{Lin:2005} For 
small ($a \ll \lambda )$ lossless dielectric particles placed in a 
standing wave of light, one may define an approximate potential energy for the 
light-induced mechanical interaction as

\begin{equation}
\label{eq4}
{\begin{array}{*{20}c}
 {U = - \sum\limits_{n = 1}^N {\left( {\alpha / 4} \right)\vert 
\mathord{\buildrel{\lower3pt\hbox{$\scriptscriptstyle\rightharpoonup$}}\over 
{E}} _{in} 
(\mathord{\buildrel{\lower3pt\hbox{$\scriptscriptstyle\rightharpoonup$}}\over 
{r}} _n )\vert ^2 - \alpha ^2 / 2} } \hfill \\
 {\times \sum\limits_{n = 1}^N {\sum\limits_{m < n} 
{\mathord{\buildrel{\lower3pt\hbox{$\scriptscriptstyle\rightharpoonup$}}\over 
{E}} _{in} 
(\mathord{\buildrel{\lower3pt\hbox{$\scriptscriptstyle\rightharpoonup$}}\over 
{r}} _m )Re\left\{ 
{\mathord{\buildrel{\lower3pt\hbox{$\scriptscriptstyle\leftrightarrow$}}\over 
{G}} 
(\mathord{\buildrel{\lower3pt\hbox{$\scriptscriptstyle\rightharpoonup$}}\over 
{r}} _n - 
\mathord{\buildrel{\lower3pt\hbox{$\scriptscriptstyle\rightharpoonup$}}\over 
{r}} _m )} \right\}} } 
\mathord{\buildrel{\lower3pt\hbox{$\scriptscriptstyle\rightharpoonup$}}\over 
{E}} _{in} 
(\mathord{\buildrel{\lower3pt\hbox{$\scriptscriptstyle\rightharpoonup$}}\over 
{r}} _n )} \hfill \\
\end{array} },
\end{equation}

\noindent
where $\alpha = 4\pi \varepsilon _0 a^3(\varepsilon - 1) / (\varepsilon + 
2)$ and 

$\mathord{\buildrel{\lower3pt\hbox{$\scriptscriptstyle\leftrightarrow$}}\over 
{G}} 
(\mathord{\buildrel{\lower3pt\hbox{$\scriptscriptstyle\rightharpoonup$}}\over 
{R}} ) = e^{ikR} / 4\pi \varepsilon _0 R^3\left[ {{\begin{array}{*{20}c}
 {( - k^2R^2 - 3ikR + 3)\hat {R}\hat {R}^T} \hfill \\
 { + (k^2R^2 + ikR - 
1)\mathord{\buildrel{\lower3pt\hbox{$\scriptscriptstyle\leftrightarrow$}}\over 
{I}} } \hfill \\
\end{array} }} \right]$. To leading orders, the 
force matrices for the three branches, evaluated using the P.E. model, are


\begin{equation}
\label{eq5}
\begin{array}{l}
 
(\mathord{\buildrel{\lower3pt\hbox{$\scriptscriptstyle\leftrightarrow$}}\over 
{K}} 
_{\mathord{\buildrel{\lower3pt\hbox{$\scriptscriptstyle\rightharpoonup$}}\over 
{k}} - branch} )_{lq} = \\ 
 \left\{ {{\begin{array}{*{20}c}
 {K_{local} (l) - \beta \sum\nolimits_{n = 1,n \ne l}^N {\left( {\vert l - 
n\vert \pi } \right)^{ - 1}} } \hfill & {\left( {l = q} \right)} \hfill \\
 {\beta \left( {\vert l - q\vert \pi } \right)^{ - 1}} \hfill & {\left( {l 
\ne q} \right)} \hfill \\
\end{array} }} \right. \\ 
 \end{array},
\end{equation}

\begin{equation}
\label{eq6}
\begin{array}{l}
 
(\mathord{\buildrel{\lower3pt\hbox{$\scriptscriptstyle\leftrightarrow$}}\over 
{K}} 
_{(\mathord{\buildrel{\lower3pt\hbox{$\scriptscriptstyle\rightharpoonup$}}\over 
{k}} \times 
\mathord{\buildrel{\lower3pt\hbox{$\scriptscriptstyle\rightharpoonup$}}\over 
{E}} ) - branch} )_{lq} = \\ 
 \left\{ {{\begin{array}{*{20}c}
 { - \beta \sum\nolimits_{n = 1,n \ne l}^N {\left[ {\begin{array}{l}
 2\left( {\vert l - n\vert \pi } \right)^{ - 3} \\ 
 - 3\left( {\vert l - n\vert \pi } \right)^{ - 5} \\ 
 \end{array}} \right]} } \hfill & {\left( {l = q} \right)} \hfill \\
 {\beta \left[ {2(\vert l - q\vert \pi )^{ - 3} - 3(\vert l - q\vert \pi )^{ 
- 5}} \right]} \hfill & {\left( {l \ne q} \right)} \hfill \\
\end{array} }} \right. \\ 
 \end{array},
\end{equation}

\noindent
and

\begin{equation}
\label{eq7}
\begin{array}{l}
 
(\mathord{\buildrel{\lower3pt\hbox{$\scriptscriptstyle\leftrightarrow$}}\over 
{K}} 
_{\mathord{\buildrel{\lower3pt\hbox{$\scriptscriptstyle\rightharpoonup$}}\over 
{E}} - branch} )_{lq} = \\ 
 \left\{ {{\begin{array}{*{20}c}
 { - \beta \sum\nolimits_{n = 1,n \ne l}^N {\left[ {\begin{array}{l}
 4\left( {\vert l - n\vert \pi } \right)^{ - 3} \\ 
 - 9\left( {\vert l - n\vert \pi } \right)^{ - 5} \\ 
 \end{array}} \right]} } \hfill & {\left( {l = q} \right)} \hfill \\
 {\beta \left[ {4(\vert l - q\vert \pi )^{ - 3} - 9(\vert l - q\vert \pi )^{ 
- 5}} \right]} \hfill & {\left( {l \ne q} \right)} \hfill \\
\end{array} }} \right. \\ 
 \end{array},
\end{equation}

\noindent
where

$K_{local} (l) = - 2k^2\alpha E_0^2 - \beta \sum\nolimits_{n = 1,n \ne l}^N 
{\left( {\vert l - n\vert \pi } \right)^{ - 1}} ,$

\noindent
$\beta = k^5\alpha ^2E_0^2 / 2\pi \varepsilon _0 ,$ and $l$ and $q$ are particle 
indices. The $I.P.R.$ computed using the P.E. model is plotted in 
Fig. \ref{fig1} as dotted lines, which are surely not quantitative compared
with the exact result, but nevertheless captures the salient features of the rigorous 
calculations.

It is evident from (\ref{eq6}) and (\ref{eq7}) that the modes of the 
$(\mathord{\buildrel{\lower3pt\hbox{$\scriptscriptstyle\rightharpoonup$}}\over 
{k}} \times 
\mathord{\buildrel{\lower3pt\hbox{$\scriptscriptstyle\rightharpoonup$}}\over 
{E}} )$-branch and the 
$\mathord{\buildrel{\lower3pt\hbox{$\scriptscriptstyle\rightharpoonup$}}\over 
{E}} $-branch are similar, because the leading terms are essentially an 
action-reaction couplings between every pair of particles, with the coupling 
strength being proportional to inverse-cubic distance. These two branches 
are more localized than those of the ball and spring model because the 
interaction has a longer range.\cite{The:1976} The 
$\mathord{\buildrel{\lower3pt\hbox{$\scriptscriptstyle\rightharpoonup$}}\over 
{k}} $-branch is the most localized and interesting. Its force matrix consists 
of two components, the long range (inverse distance) action-reaction 
coupling and $K_{local} (l)$ which acts like a spring that 
ties the \textit{l-th} particle to its equilibrium position. The first term of $K_{local} 
(l)$ is caused by the incident beam and is the same for each particle. This 
term gives a frequency gap at low frequency (e.g. between 0 and 4.7 MHz in 
Fig. \ref{fig1}(a)), while the second term is induced by OB. 
One may define an intrinsic vibration frequency for every individual 
particle as

\begin{equation}
\label{eq9}
\Omega _{\mbox{intrinsic}} (l) = \sqrt { - K_{local} (l) / m} ,
\end{equation}

\noindent
plotted in Fig. \ref{fig2}(g). We note that the first term of 
$K_{local} (l)$ contributes a constant to $\Omega _{\mbox{intrinsic}} (l)$, 
while the term due to OB gives a position dependent contribution that makes 
$\Omega _{\mbox{intrinsic}} (l)$ higher (lower) near the center (ends) of 
the lattice. It is the variation of $\Omega _{\mbox{intrinsic}} (l)$ along 
the chain that elicits the enhanced localization: only particles near the 
center (both ends) participate in the high (low) frequency vibrations, see 
Fig. \ref{fig2}(c) (Fig. \ref{fig2}(f)).

We now consider the strength of the OB. As revealed by recent 
theoretical\cite{Lin:2005}$^{,}$\cite{Antonoyiannakis:1997} and 
experimental\cite{Burns:1990}$^{,}$\cite{Tatarkova:2002}$^{,}$\cite{Singer:2003}$^{,}$\cite{Garces:2005}$^{,}$\cite{Mellor:2006} 
works, the optical force on microspheres can dominate over other relevant 
interactions such as gravity, van der Waals, and thermal fluctuations. For 
the lattice consists of smaller spheres defined in (\ref{eq2}), the potential energy 
per particle $U / N$ for $N$=1, 10, 50, 100 are respectively -9.6, -10.5, 
-11.3, -11.7 $k_{B}T_{Room}$, and the chain should thus be thermally stable 
at the assumed intensity. Furthermore, $U / N$ is enhanced by more than 
20{\%} as $N$ is increased from 1 to 100, implying that the OB carries a 
non-negligible contribution.

We have showed that OB can bind a collection of particles into a 1D lattice 
that is stable in all three dimensions. We shall emphasize that the 
localization discussed here is a general phenomena for optically-bound 
structures that are spatially extended, and it is not restricted to the 
particular geometry or incident wave considered here. We found that LMs are 
also observed in other structures such as the photonic cluster made from 
microspheres shown in Fig. 4(\textbf{g}) of reference 
\onlinecite{Lin:2005}, and also another lattice configuration defined by 
${\begin{array}{*{20}c}
 
{\mathord{\buildrel{\lower3pt\hbox{$\scriptscriptstyle\rightharpoonup$}}\over 
{R}} _n = \left( {0,n\lambda ,0} \right),} \hfill & {n = 1,2\ldots ,N} 
\hfill \\
\end{array} }.$ A difference between this lattice configuration and (\ref{eq2}) is 
that the lattice constant of the later (former) is dictated by optical 
binding (trapping). It is the long-ranged OB that induces the variation 
of $\Omega _{\mbox{intrinsic}} (l)$, which in turn induce the localization.

It is worth to note that in the case of the 1D array specified by (\ref{eq2}), the 
stable configuration is defined by optical traps produced by the incident 
wave rather than OB, yet OB plays a crucial role on the dynamics. The 
quasi-stable dynamics that arises from the nonconservative nature of the 
optical forces,\cite{Lin:2005} and the LMs considered here, 
could be major causes of the inconsistencies between the vast amount of 
light-trapping experiments and theoretical predictions where OB is 
neglected.\cite{See:2003} A deeper investigation into the subject 
would be an interesting and important research topic for the future.

Support by CA02/03.SC05 is gratefully acknowledged. We thank Kin-Hung Fung 
for useful discussions. C.T. Chan's e-mail address is phchan@ust.hk.

\begin{thebibliography}{16}
\bibitem{Ashkin:1970} A. Ashkin, Phys. Rev. Lett. \textbf{24}, 156 (1970).
\bibitem{See:2003} See e.g. D.G. Grier, Nature \textbf{424}, 810 (2003).
\bibitem{Burns:1990} M.M. Burns, J.M. Fournier, and J.A. Golovchenko, Science \textbf{249}, 749 (1990).
\bibitem{Lin:2005} J. Ng, Z.F. Lin, C.T. Chan, and P. Sheng, Phys. Rev. B \textbf{72}, 085130 (2005); \textit{ibid}, \textit{Opt. Lett.} \textbf{30}, 1956 (2005).
\bibitem{Chaumet:2001} P.C. Chaumet and M. Nieto-Vesperinas, Phys. Rev. B \textbf{64}, 035422 (2001).
\bibitem{Campbell:2004} D.K. Campbell, S. Flach, and Y.S. Kivshar, Physics Today \textbf{57}, \textit{No.} 1, 43 (2004).
\bibitem{Tatarkova:2002} S.A. Tatarkova, A.E. Carruthers, and K. Dholakia, Phys. Rev. Lett. \textbf{89}, 283901 (2002).
\bibitem{Singer:2003} W. Singer, M. Frick, S. Bernet, and M. Ritsch-Marte, J. Opt. Soc. Am. B \textbf{20}, 1568 (2003).
\bibitem{Black:2003} A.T. Black, Hilton, W. Chan, and V. Vuletic, Phys. Rev. Lett. \textbf{91}, 203001 (2003).
\bibitem{Garces:2005} V. Garces-Chavez and K. Dholakia, Appl. Phys. Lett. \textbf{86}, 031106 (2005).
\bibitem{Mellor:2006} C.D. Mellor and C.D. Bain, ChemPhysChem \textbf{7}, 329 (2006).
\bibitem{The:1} The stable configurations calculated by the MS-MST formalism deviate from (\ref{eq2}) by less than 0.003$\lambda$.
\bibitem{This:1} This is so because, on every sphere, the path difference between the incident field and the scattered field from the other spheres, are roughly an integer multiple of 2$\pi $, which enhances the stability.
\bibitem{Chowdhury:1985} A. Chowdhury and B. Ackerson, Phys. Rev. Lett. \textbf{55}, 833 (1985).
\bibitem{Book:1} N.E. Cusack, The Physics of Structurally Disordered Matter: An Introduction (A. Hilger, Philadelphia, 1987), p. 239.
\bibitem{The:1976} The finite coherent length of real laser will effectively set an upper limit on \textit{N}. In the hypocritical case where $N \to \infty $, the modes for (\ref{eq6})-(\ref{eq7}) become extended modes, whereas (\ref{eq5}) diverges.
\bibitem{Antonoyiannakis:1997} M.I. Antonoyiannakis and J.B. Pendry, Phys. Rev. B \textbf{60}, 613 (1997).
\end{thebibliography}
\end{document}